\documentclass[letterpaper, 10 pt, conference]{ieeeconf}

\IEEEoverridecommandlockouts

\usepackage{marvosym} 
\usepackage{booktabs} 
\usepackage{graphicx}
\usepackage{multirow}
\usepackage{threeparttable}
\usepackage{balance}

\usepackage[hidelinks]{hyperref}

\usepackage{siunitx}
\usepackage{xcolor}

\title{\LARGE \bf
Computationally efficient neural network classifiers for next generation closed loop neuromodulation therapy -- a case study in epilepsy
}

\author{
Ali Kavoosi$^{1 \, \dag  \,\text{\normalsize \Letter}}$, Robert Toth$^{2 \, \dag}$, Moaad Benjaber$^{1}$, Mayela Zamora$^{1}$,\\ Antonio Valent{\'i}n$^{3}$, Andrew Sharott$^{2}$ and Timothy Denison$^{1,2 \, \text{\normalsize \Letter}}$
\thanks{This work was supported by the John Fell Fund of the University of Oxford, the UK Medical Research Council (MC\_UU\_00003/3, MC\_UU\_00003/6) and the Royal Academy of Engineering.}%
\thanks{$\dag$ Ali Kavoosi and Robert Toth contributed equally to this work and share first authorship.}%
\thanks{$^{1}$ Brain Network Dynamics Unit, Department of Pharmacology, University of Oxford, Oxford OX1~3TH, United Kingdom}%
\thanks{$^{2}$ Institute of Biomedical Engineering, Old Road Campus Research Building, Department of Engineering Sciences, University of Oxford, Oxford OX3~7DQ, United Kingdom}%
\thanks{$^{3}$ Department of Basic and Clinical Neuroscience, King's College London, London SE5 9RT, United Kingdom}%
\thanks{$^{\text{\normalsize \Letter}}$ {ali.kavoosi@linacre.ox.ac.uk, timothy.denison@eng.ox.ac.uk}}%
}

\usepackage[pscoord]{eso-pic}
\newcommand{\placefootnote}[1]{
  \AddToShipoutPictureFG*{
  \put(\LenToUnit{0.7in}, \LenToUnit{0.7in-1.35\baselineskip}){
    \begin{minipage}{\textwidth}
    \centering
    {\footnotesize
    This paper was presented at the 44th Annual International Conference of the IEEE Engineering and Medicine in Biology Society, July 11--15th 2022, Glasgow, Scotland, United Kingdom. The published version is available from IEEE Xplore at https://doi.org/10.1109/EMBC48229.2022.9871793.}
    \end{minipage}
   }
  }
}

\placefootnote

\begin{document}

\bstctlcite{IEEE:BSTcontrol}

\maketitle
\thispagestyle{empty}
\pagestyle{empty}

\begin{abstract}This work explores the potential utility of neural network classifiers for real-time classification of field-potential based biomarkers in next-generation responsive neuromodulation systems. Compared to classical filter-based classifiers, neural networks offer an ease of patient-specific parameter tuning, promising to reduce the burden of programming on clinicians. The paper explores a compact, feed-forward neural network architecture of only dozens of units for seizure-state classification in refractory epilepsy. The proposed classifier offers comparable accuracy to filter-classifiers on clinician-labeled data, while reducing detection latency. As a trade-off to classical methods, the paper focuses on keeping the complexity of the architecture minimal, to accommodate the on-board computational constraints of implantable pulse generator systems.
\newline

\indent \textit{Clinical relevance}---A neural network-based classifier is presented for responsive neurostimulation, with comparable accuracy to classical methods at reduced latency. 
\end{abstract}

\section{INTRODUCTION}
Deep brain stimulation (DBS) first received approval for the symptomatic treatment of Parkinson's disease in 1997. While similar in design to cardiac pacemakers, the implantable pulse generators (IPG) of the time offered only an open-loop form of therapy, with typically a single stimulation pattern, set by a clinician for each patient. Real-time seizure detection and responsive neurostimulation (RNS) was first attempted using a computer-in-the-loop system by Gotman et al. in 1976 \cite{Gotman_1976}, it was Osorio et al. in 1998 \cite{Osorio_1998, Osorio_2002, Bhavara_2006} who introduced the more widely studied filter-based spectral biomarker detectors to the field of epilepsy research. With the continued development of IPGs and the maturation of low-power microprocessor technology, the first RNS system for epilepsy received approval for pre-clinical use in 2014. This system from Neuropace had the capabilities to sense bioelectric signals, and choose stimulation programs based on clinician-configured classification state \cite{Sun_2014}.

Filter-based spectral-feature detectors have since been used successfully in other conditions, most notably for tremor suppression in Parkinson's disease, through the discovery of beta oscillations as a correlate of disease state \cite{Little_2013}. However, the smaller ($\SI{1}{\micro\volt rms}$) signal size of beta oscillations, compared to epileptiform activity ($\SI{10}{\micro\volt rms}$), made deploying the detector algorithm in IPGs challenging due to the presence of stimulation and other artifacts. Contemporary work focuses on improving the robustness of the signal chains to enable simultaneous sensing and stimulation, thus true closed-loop operation across targeted diseases \cite{Sorkhabi_2020, Anso_2022}. Examples include the Medtronic Percept \cite{Neumann_2021} and the Picostim-DyNeuMo research systems \cite{Zamora_2022, Toth_2020}. A complementary avenue of refinement is the use of feed-forward predictors for adapting stimulation based on periodicities of disease state and patient needs, such the circadian scheduling of the SenTiva system from LivaNova or the Picostim-DyNeuMo \cite{Zamora_2021}. Taking advantage of more long term, weekly or even monthly rhythms are being investigated for epilepsy management \cite{Baud_2018}.

Patient-specific filter design, while possible to aid with software, can be a complex problem, likely to limit both clinician-capacity and patient-throughput. Establishing and validating a neural-network (NN) training pipeline based on clinician-labeled data could offer a systematic classifier tuning process. Networks could be pre-trained on aggregate data from multiple patients, and refined based on individual labeled data at the point of deployment \cite{LeCun_2015}. Of course, due to the black-box nature of neural network classifiers, extensive validation work will be required to establish safety before first-in-human studies. Advances in interpretable deep learning could facilitate building trust in NN--classifiers for medical use \cite{Lapuschkin_2019}.

Liu et al. \cite{Liu_2021} demonstrated the feasibility of deploying high accuracy classifiers for seizure detection on modern microprocessors (ARM\textsuperscript{\tiny\textregistered} Cortex-M4), through model compression and quantization techniques, showcasing several advanced NN topologies.

This paper is meant as an initial study to bring focus to the fundamental challenge of NN classifiers: computational cost. As state-of-the-art deep neural networks reach ever increasing model sizes \cite{LeCun_2015, Bender_2021}, we aim to explore whether lean NNs of only dozens of units could in fact compete in accuracy with classical, filter-based systems for bioelectric signal classification.

\begin{figure*}[!t]
    \centering
    \includegraphics[width=160mm]{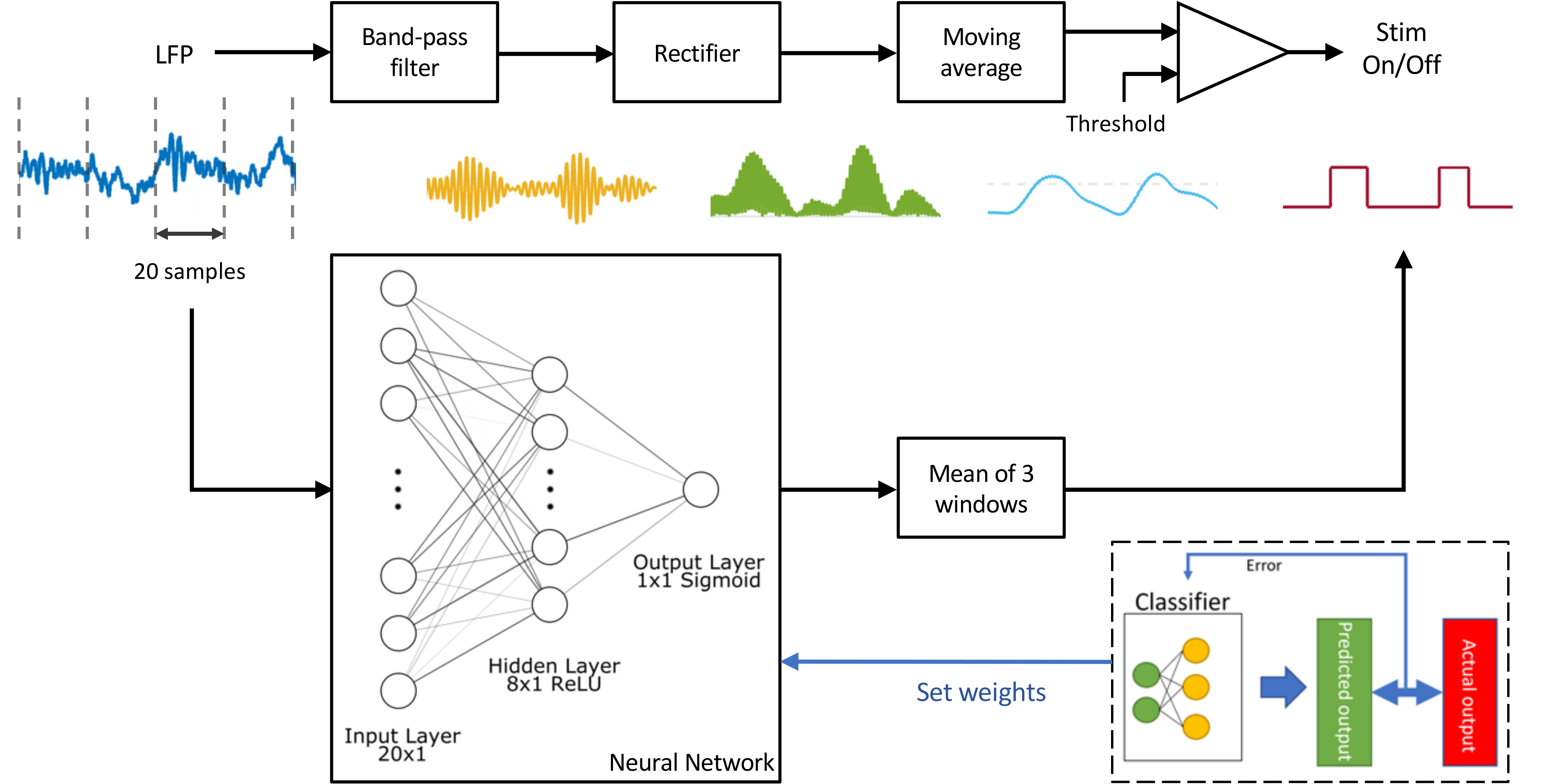}
    \caption{Classifier architectures. {\it Top:} classical filter-based spectral power detector \cite{Little_2013}. {\it Bottom:} the multi-layer perceptron architecture evaluated in this paper. Through training on labeled data, the neural network is expected to assume an overall transfer function similar to the hand-crafted filter topology.}
    \label{fig:architecture}
    \vspace{-1.0\baselineskip}
\end{figure*}

\section{Design}

\subsection{Baseline Method} 
To establish a baseline for performance as well as for computational cost, we used a classical band-power estimation filter chain to detect epileptiform discharges \cite{Osorio_2002, Little_2013}, that we have previously deployed with success in the Picostim-DyNeuMo experimental IPG system \cite{Toth_2020}. The processing steps of this method are shown in the top panel of Fig.\,\ref{fig:architecture}. While this algorithm is computationally efficient and has a very favorable memory footprint (refer to Table~\ref{tab:resources}), the demodulated envelope signal, thus the detector output, will always lag the input signal to reduce output ripple -- irrespective of processing speed. This trade-off arises from the very nature of causal filtering, and is necessary to prevent rapid switching of the detector output for input signals near the classification threshold. The reference classifier was configured as follows. Our band-pass stage was an $8-22$\,Hz, 4th order Butterworth filter, with a Direct Form I IIR implementation (16-bit coefficients, 32-bit accumulators). Envelope demodulation was achieved using an exponential moving average filter with a decay-coefficient of 32 samples. The filter chain, and all other classifiers were designed to operate at a sampling rate of 256\,Hz.

\subsection{New neural-based methods}
In our search for a low-complexity classifier for time series input, we explored two main NN families. (1) Multi-Layer Perceptrons (MLP) are the simplest, and oldest family or of artificial neural networks \cite{Rosenblatt_1958}, where the input vector is connected to `hidden' layers of feedforward units, condensing information into an output unit. This architecture is shown in the bottom panel of Fig.\,\ref{fig:architecture}. (2) As a step up in complexity, Convolutional Neural Network (CNN) introduce a convolution layer, also known as a filter bank, between the input vector and the neural layers as an initial extra step \cite{LeCun_2015}. The input to our networks is  formed by a windowed set of past time samples of the local field potential (LFP) signal. The output signal, calculated once for each complete window of samples, is thresholded into a binary label. We denote this classifier the `standalone MLP' model.

Recurrent neural networks, an otherwise natural choice for processing time series data, were dismissed from consideration as recurrence necessitates the introduction of dynamic state variables, which significantly increases memory footprint \cite{LeCun_2015}. Without recurrence, we introduced coherence into our classifier in a different way. We settled on requiring a consensus of three subsequent outputs from the NN to define our final output label, providing the `adjusted MLP' model.

\subsection{Training and data}
Our raw dataset consisted of LFP signals recorded from two patients, for a combined 24 hours, with 30 professionally labeled events of clinical significance. The recordings were resampled to a 256\,Hz sampling frequency for uniformity.

As seizures are comparatively rare events scattered among very long periods of normal activity, we decided to introduce class imbalance into our training sets to best prepare the NNs for real-life use. The training set was biased towards negative samples in a 3:1 ratio, based on clinician annotations. The dataset was split in the common 70:30 ratio between a training and a validation set. Network weights and biases were quantized to 8-bit integers.

\subsection{Technical equipment used}
Neural networks were modeled and trained in Tensorflow Lite version 2.7.0, using an Intel Core i7 CPU with 16 GB of RAM. Embedded performance was tested on an Arduino Nano 33 BLE Sense evaluation board for the nRF52840 ARM Cortex-M4F microprocessor.

\begin{figure}[!t]
    \centering
    \includegraphics[width=8.6cm]{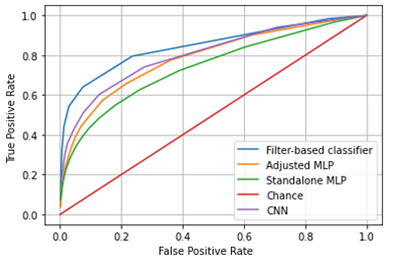}
    \vspace{-2.0\baselineskip}
    \caption{ROCs for different classifiers. Note that performance converges towards high TPR and FPR, which is the desirable operating point of seizure detectors as FNs pose significantly greater risk of harm to patients than FPs.}
    \label{fig:roc}
    \vspace{-1.0\baselineskip}
\end{figure}

\subsection{Comparison of CNN and MLP models}
Fig.\,\ref{fig:roc} shows the performance of our CNN and MLP classifiers. While the CNN outperforms the standalone MLP, it performs with similar accuracy to the adjusted MLP model above 80\% true positive rate (TPR). For safe use, the operating point of a seizure detection system should be biased towards high TPR -- missed seizures (false negatives) pose significantly more risk to the patient than false positives, which merely result in unnecessary stimulation. Overall, in targeting resource constrained IPGs, we judged the minor edge of the CNN insufficient to justify the added computational burden of the convolutional layer.

\subsection{Tuning the MLP classifier}
\begin{figure}[!t]
    \centering
    \includegraphics[width=86mm]{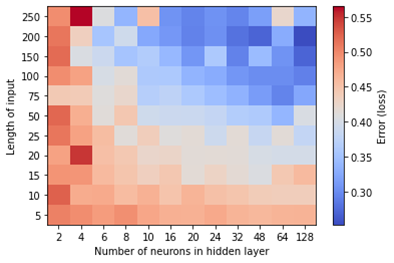}
    \vspace{-2.0\baselineskip}
    \caption{Tuning the MLP classifier. Grid search on the two main model hyperparameters: the input vector length and the number of hidden layer units. Loss is represented as binary cross entropy across all samples.}
    \label{fig:hyperparameters}
    \vspace{-1.0\baselineskip}
\end{figure}

The performance of the MLP model, when trained on a given dataset, is primarily determined by two hyperparameters: the number of timepoints in the input window, and the hidden layer’s size. We found that varying the number of hidden layers had very modest effects on accuracy (not shown in this paper). Fig.\,\ref{fig:hyperparameters} systematically explores the effect of the two key hyperparameters on the classification error of a single output, single hidden layer MLP model. As expected, the network requires a certain size and complexity to encode a feature space sufficient for reliable classification, though increasing the number of units in either layer beyond a certain point leads to diminishing returns. To select one of the possible models from the error surface, one could define a scoring scheme including network size, computational time and the loss itself, to make an educated choice, however, this is beyond the scope of this paper. Favoring low complexity, we settled on using a 20-point input window and 8 hidden neurons, in the `transition zone' of the error surface.

\begin{figure}[!b]
    \vspace{-1.0\baselineskip}
    \centering
    \includegraphics[width=8.6cm]{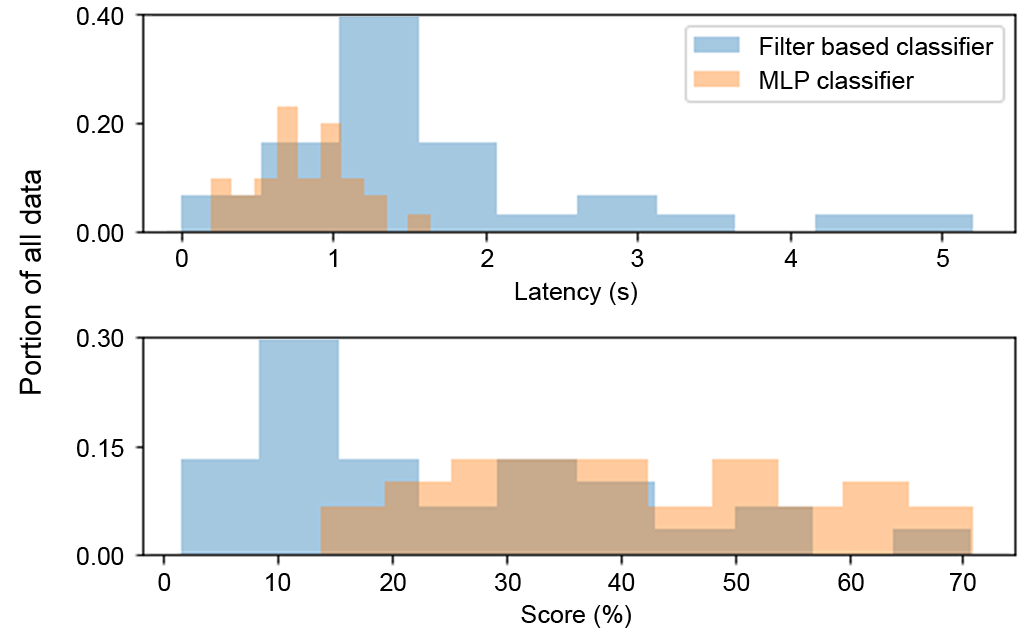}
    \vspace{-2.0\baselineskip}
    \caption{Detailed performance of classifiers. {\it Top:} histogram of classification latency. {\it Bottom:} histogram showing the percentage of overlap between positive classifier output and clinician-labeled event.}
    \label{fig:hist}
\end{figure}

\section{Model Performance and Interpretation}

The next step is to compare our best MLP result to the baseline classifier. The ROC of Fig.\,\ref{fig:roc} reveals that a well-tuned filter chain outperforms the small MLP model below 60\% false positive rate, beyond which they converge in accuracy. Identifying a seizure does not present a holistic view of performance though. In Fig.\,\ref{fig:hist} we highlight two additional characteristics to consider in classifier evaluation: (1) latency at event onset, and (2) the overlap between classifier and clinician labels. As shown, the MLP responds on average more rapidly to a commencing seizure (mean latency of 0.6\,sec vs 1.7\,sec), and tracks the clinician label more closely overall, compared to the baseline method.

\begin{figure*}[!t]
    \centering
    \includegraphics[width=170mm]{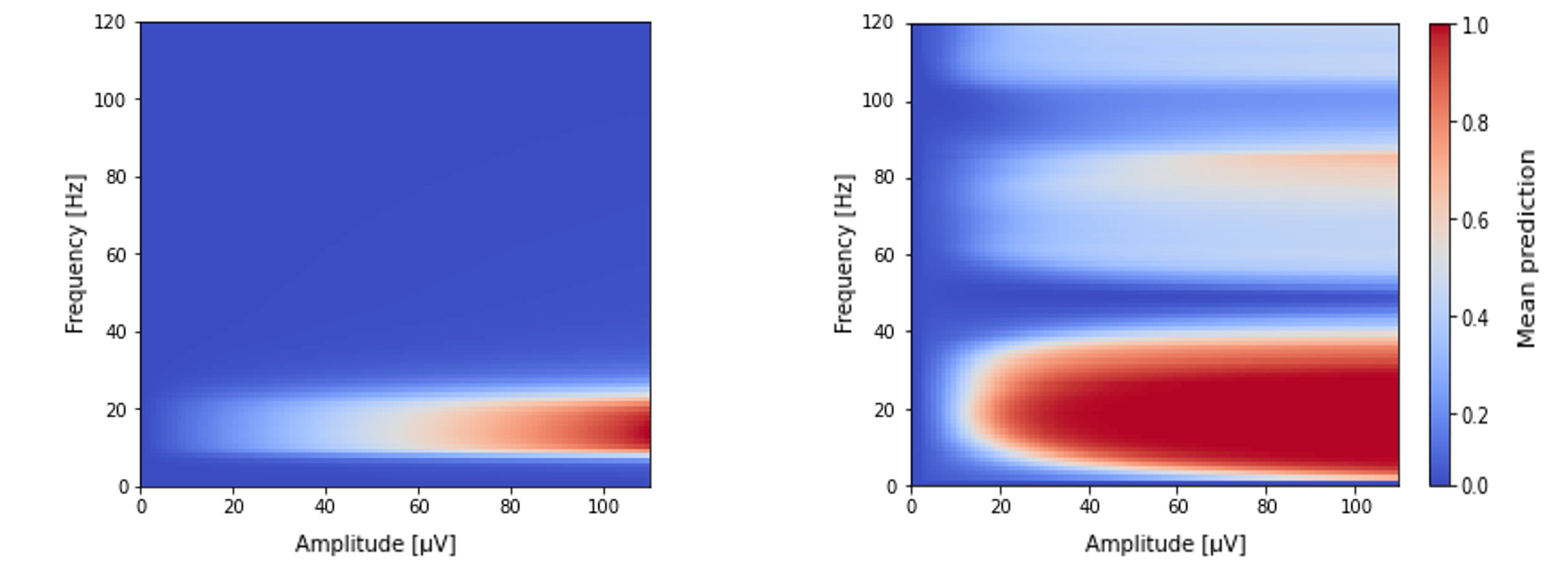}
    \vspace{-1.0\baselineskip}
    \caption{Frequency response of classifiers to input signals of different magnitudes. {\it Left:} filter chain classifier. {\it Right:} MLP classifier with average of 3 windows. Accuracy is presented as the mean classifier output for a 1\,second sinusoidal test tone over 10 repeats.}
    \label{fig:interpretation}
\end{figure*}

To explore the MLP classifier's internal representation of a seizure, we present a small interpretation experiment in Fig.\,\ref{fig:interpretation}. We presented the classifiers with second-long sinusoidal bursts of activity, performing parameter sweeps along both test frequency and test amplitudes. As seen, the MLP model (right) was successful in internalizing a notion of the spectral characteristics of epileptiform activity (low frequency lobe), that encompasses the pass-band of the filter classifier (left). The greater effective bandwidth of the MLP could explain the lower false positive rates seen on the ROC of Fig.\,\ref{fig:roc}. The activation lobes at higher frequencies likely represent a process analogous to aliasing, and we expect this periodicity to be a correlate of the input window size, which should be investigated further.

\begin{table}[!ht]
\centering
\begin{threeparttable}
    \caption{Resource Usage}
	\label{tab:resources}
	\centering
	\begin{tabular}{c >{\centering\arraybackslash}p{20mm} c c}
		\toprule[\lightrulewidth]
		\multirow{2}{*}{Classifier} & Execution time & Code & Memory \\
		& (Cycles\,/\,Sample) & (bytes) & (bytes) \\
		\midrule
		Filter (custom) & 750 & 675 & 100 \\
		MLP (TF Lite) & 2250 & 31k & 8k \\
		CNN (TF Lite) & 3200 & 31K & 13k \\
		MLP (custom) & \textcolor{white}{*}2250* & \phantom{*}2.0k* & \phantom{*}300* \\
		\bottomrule[\lightrulewidth]
	\end{tabular}
    \begin{tablenotes}
      \small
      \item *predicted
    \end{tablenotes}
\end{threeparttable}
\vspace{-1.0\baselineskip}
\end{table}

\section{DISCUSSION}
The example design of the MLP classifier demonstrates that even tiny neural networks can be effective at simple signal processing tasks. As the final step, we should reflect on the embedded resource usage achieved, so we refer the reader to Table~\ref{tab:resources}. Importantly, the network achieved sufficiently low complexity for real time use. Note that NN execution times are reported per sample, though the output only changes at the end of a window of 20 samples. Notably, the true memory footprint of the classifier could not be determined with this evaluation system -- Tensorflow Lite does not generate network code, rather it provides a network description file, to be run by a relatively large, general purpose interpreter library in the embedded system. For a more realistic, yet conservative outlook, we present estimates for the memory usage of the same network deployed using customized library, trimmed down to the features used in our design. In summary, the MLP could provide an alternative to existing tuned filter methods used in commercial devices.




\section*{DISCLOSURES}

The University of Oxford has research agreements with Bioinduction Ltd. Tim Denison also has business relationships with Bioinduction for research tool design and deployment, and stock ownership ($<\SI{1}{\percent}$).

\section*{ACKNOWLEDGMENT}

The authors would like to thank Bence Mark Halpern for reviewing the manuscript, and Tom Gillbe at Bioinduction Ltd. for feedback on the filter-classifier design.

\balance

\vspace{1.0\baselineskip}




\end{document}